# Quasi two-dimensional nature of high-$T_c$ superconductivity in iron-based (Li,Fe)OHFeSe

Dong Li[1,†,*], Yue Liu[1,2,†], Zouyouwei Lu[1,2,†], Peiling Li[1], Yuhang Zhang[1,2], Sheng Ma[1,2], Jiali Liu[1,2], Jihu Lu[1,2], Hua Zhang[1,2,3], Guangtong Liu[1,2,3], Fang Zhou[1,2,3], Xiaoli Dong[1,2,3,4*], and Zhongxian Zhao[1,2,3,4]

[1] *Beijing National Laboratory for Condensed Matter Physics, Institute of Physics, Chinese Academy of Sciences, Beijing 100190, People's Republic of China.*
[2] *School of Physical Sciences, University of Chinese Academy of Sciences, Beijing 100049, People's Republic of China.*
[3] *Songshan Lake Materials Laboratory, Dongguan, Guangdong 523808, People's Republic of China.*
[4] *Key Laboratory for Vacuum Physics, University of Chinese Academy of Sciences, Beijing 100049, People's Republic of China.*

† These authors contributed equally to this work.
* Correspondence to: lidong@iphy.ac.cn and dong@iphy.ac.cn;

**The intercalated iron selenide (Li,Fe)OHFeSe has a strongly layered structure analogous to the quasi two-dimensional (2D) bismuth cuprate superconductors, and exhibits both high-temperature ($T_c$) and topological superconductivity. However, the issue of its superconductivity dimensionality has not yet been fully investigated so far. Here we report that the quasi-2D superconductivity features, including the high anisotropy $\gamma = 151$ and the associated quasi-2D vortices, are also revealed for (Li,Fe)OHFeSe, based on systematic experiments of the electrical transport and magnetization and model fittings. Thus, we establish a new vortex phase diagram for (Li,Fe)OHFeSe, which delineates an emergent quasi-2D vortex-liquid state, and a subsequent vortex-solid dimensional crossover from a pancake-like to a three-dimensional state with decreasing temperature and magnetic field. Furthermore, we find that all the quasi-2D characteristics revealed here for the high-$T_c$ iron selenide superconductor are very similar to those reported for the high-$T_c$ bismuth cuprate superconductors.**

**PACS**:74.70.Xa;74.78.-w;74.25.F-;74.25.Uv

**Introduction**

The issue of the two-dimensionality effect[1] is one of the central concerns in understanding the high-$T_c$ superconductivity that emerges from the strongly layered compounds of the iron-based and cuprate families. It has been suggested that the cooper pairs are pre-formed well above the superconducting (SC) transition $T_c$ in, for instance, the molecule-intercalated iron selenides[2,3] and layered 12442-type iron arsenides $A$Ca$_2$Fe$_4$As$_4$F$_2$ ($A$ = K and Cs),[4,5] due to the two-dimensionality-induced strong phase fluctuations responsible for the pseudogap within the pre-pairing scenario.[1] With the advance in experimental techniques, two-dimensional (2D) superconductors such as high-$T_c$ monolayer FeSe on SrTiO$_3$ substrate[6-8] can now be prepared by various methods,[9-12] with the sample thickness being smaller than the $c$-axis coherence length $\xi_c$. The 2D phase fluctuations[13] are also proposed to understand the high gap opening temperature over 65 K[6,14] observed in the monolayer FeSe/SrTiO$_3$ superconductor. Although abundant novel properties have been observed in the 2D superconductors,[15-21] controversy remains about the two-dimensionality effect because the superconducting phase coherence should be three-dimensional (3D) to hold the long-range order.[22] The interlayer coupling still plays an important role in the superconductivity of the intercalated iron selenides[23,24] and layered 12442 iron arsenides.[25] Therefore, clarifying the interlayer-coupling nature appears to be important for an understanding of the highly anisotropic superconductivity arising in strongly layered iron-based and cuprate compounds.

For the highly anisotropic cuprate superconductors of Bi$_2$Sr$_2$CaCu$_2$O$_{8+\delta}$ (Bi-2212, $T_c$ ~ 84 K), previous studies[26,27] have shown that the applied $c$-axis magnetic field is conducive to the appearance of the quasi-2D feature. When the $c$-axis field is increased to a dimensional crossover field $H_{2D}$ at a given temperature,[28,29] it starts to significantly disturb the interlayer vortex coupling between the superconducting layers, changing the 3D vortex structure into a quasi-2D pancake-like one.[30] The interlayer coupling between the quasi-2D pancake vortices has proved to be a Josephson-type one,[31] and the motion of the quasi-2D vortices in the different SC layers is independent of one another.[28] This highly anisotropic superconductivity featuring the quasi-2D vortices is distinct from the 3D anisotropic superconductivity, while closer to the 2D superconductivity. However, such a quasi-2D nature of high-$T_c$ superconductivity has not yet been reported in iron-based superconductors up to date.

Therefore, it is highly desirable to further investigate the interlayer-coupling nature in a strongly layered high-$T_c$ iron-based superconductor. In this study, we choose the intercalated iron selenide (Li,Fe)OHFeSe system,[32] showing both high-$T_c$ and topological[23] superconductivity, as a superior platform to investigate the two-dimensionality effect on the superconductivity, because of its notable quasi-2D character[33,34] and sample availability. Weak interlayer hydrogen bondings and a large interlayer spacing $d \sim 9.3$ Å are the characteristic of the (Li,Fe)OH-molecules intercalated high-$T_c$ (~42 K) (Li,Fe)OHFeSe superconductor, compared to the prototypical low-$T_c$ (~8.5 K) bulk FeSe superconductor with a much smaller $d \sim 5.5$ Å [Fig. 1(a)]. Correspondingly, (Li,Fe)OHFeSe has a quasi-2D electronic structure [34,35] similar to the 2D monolayer FeSe/SrTiO$_3$. Moreover, the recent-year development of the hydrothermal synthesis methods[36,37] for (Li,Fe)OHFeSe single crystals[33] and epitaxial films[38-41] enables us to measure the intrinsic properties on the high-quality samples.

It turns out that the high-$T_c$ intercalated (Li,Fe)OHFeSe indeed displays the highly anisotropic superconductivity features, which are revealed here by systematically measuring and fitting the electrical transport and magnetization properties, as well as by comparing the results with those obtained in the low-$T_c$ prototypical FeSe and reported in the highly anisotropic high-$T_c$ Bi-2212 cuprates. The main results are sketched as follows.

(1) First, the angular-dependent upper critical field, $H_{c2}(\theta)$, is found to conform well to the 2D Tinkham model with a well fitted high anisotropy $\gamma = 151$ in the intercalated (Li,Fe)OHFeSe, while $H_{c2}(\theta)$ of the prototypical FeSe follows the 3D anisotropic Ginzburg-Landau (G-L) model as expected.

(2) Moreover, a crossover from the quasi-2D to 3D vortex solid is determined at a threshold $H_{2D}(T)$ for (Li,Fe)OHFeSe by the magnetic relaxation experiments. The experimental crossover field $H_{2D}$, close to 1 kOe at the lowest measuring $T = 2$ K, matches well with the theoretical prediction given by $H_{2D}^{0K} \sim \Phi_0/(\gamma d)^2$ (here $\Phi_0$ is the flux quantum).[28]

(3) Furthermore, a Berezinskii-Kosterlitz-Thouless (BKT)–like transition is observed in (Li,Fe)OHFeSe by both scaling the resistance data and power law fitting the voltage-current curves. This also indicates the quasi-2D superconductivity even at zero external magnetic field.

(4) Thus, a new $H$-$T$ vortex phase diagram is established for (Li,Fe)OHFeSe. This phase diagram depicts the emergent quasi-2D vortex-liquid state, and the subsequent vortex-solid dimensional crossover at $H_{2D}(T)$ from the quasi-2D pancake to the 3D vortex state with decreasing temperature and $c$-axis magnetic field. Based on this phase diagram and previous results, we highlight that the interlayer coupling between the quasi-2D vortices plays a significant role in the highly anisotropic high-$T_c$ and topological superconductivity of (Li,Fe)OHFeSe.

(5) Finally, it is also important to find that all the quasi-2D features presently revealed in the strongly layered iron selenide superconductor are very similar to those previously reported in the highly anisotropic bismuth cuprate superconductors. This sheds a new light on the common quasi-2D physics for the iron-based and copper-based high-$T_c$ superconductivity.

## Experiments

The (Li,Fe)OHFeSe single crystals were synthesized by the hydrothermal ion-exchange method,[33] and the FeSe single crystals were grown by the chemical vapor transport method.[42] The (Li,Fe)OHFeSe epitaxial films were grown on LaAlO$_3$ substrates via the matrix-assisted hydrothermal epitaxy (MHE) methods as reported elsewhere.[38] X-ray diffraction (XRD) data of all the studied samples were collected at room temperature on a 9 kW Rigaku SmartLab x-ray diffractometer equipped with two Ge(220) monochromators. The measured samples have thicknesses of 600-700 nm for the films (determined on a Hitachi SU5000 scanning electron microscope) and of about 20 microns for the single crystals, with the lateral size of millimeters for both the samples. The electrical transport properties were measured on a Quantum Design PPMS-9 system with the standard four-probe method. During the angle-dependent measurements, the samples were rotated to change the tilting angle $\theta$ between the field $H$ and crystallographic $c$ axis. The upper limit of applied current is 5 mA in the measurements of voltage-current ($V$-$I$) curves. The magnetization measurements were performed on a Quantum Design MPMS-3 system under the $c$-axis magnetic fields. In the magnetic relaxation measurements, the $c$-axis magnetic field was swept to a field at least 10 kOe higher than the target field to ensure the full penetration of the field, as proposed in the review paper.[43] The normalized magnetic relaxation rate, $S = |d \ln M / d \ln t|$, was obtained by measuring the decay of magnetization $M$ with time $t$ for more than one hour, with $t$ counted from the initial critical state.

## Main results

Figure 1(a) schematically illustrates the strongly layered crystal structure of (Li,Fe)OHFeSe with a large interlayer spacing $d \sim 9.3$ Å between the SC FeSe layers, which is derived from the prototypical FeSe structure with $d \sim 5.5$ Å by intercalating (Li,Fe)OH molecules between the FeSe layers. All the XRD $\theta$-$2\theta$ scans of the FeSe single crystals and (Li,Fe)OHFeSe epitaxial films exhibit a single (00$l$) orientation and there are no detectable impurity phases [Figs. 1 (b-c)]. The left shift of the (00$l$) peak positions demonstrates the expanded $c$-axis lattice parameter of (Li,Fe)OHFeSe [Fig. 1 (c)] compared to that of bulk FeSe [Fig. 1 (b)] due to the molecules intercalation. The full width at half maximum (FWHM) of the XRD rocking curves is as small as 0.12° [Figs. 1 (d-e)], indicating an excellent out-of-plane lattice alignment of the studied samples.[38-40] The superconductivity was checked by the electrical transport measurements [Figs. 1 (f-g)], and $T_c$ was determined from the onset temperature of zero resistance at zero field. The $T_c$ increases from 8.5 K in FeSe to 42.0 K in (Li,Fe)OHFeSe, presumably partly due to the charge transfer from the (Li,Fe)OH interlayers.[32,44]

The dimensionality of superconductivity can be investigated by measuring the angular dependence of upper critical field, $H_{c2}(\theta)$.[45] In Figs. 2 (a) and (b), we show the magnetic-field-dependent resistance measured at different tilting angles from $\theta = 0°(H//c)$ to $\theta = 90°(H//ab)$ for the FeSe single crystals and (Li,Fe)OHFeSe epitaxial films, respectively. Their $H_{c2}$ values are extracted from the 10 % $R_n$ (the normal-state resistance near $T_c$), as indicated by the horizontal dashed lines in Figs. 2(a) and (b). The $H_{c2}$ data are plotted as functions of the tilting angle $\theta$ in Figs. 2 (c) and (d) for FeSe and (Li,Fe)OHFeSe, respectively. We have tried to fit the line shapes of their $H_{c2}(\theta)$ by both the 3D anisotropic G-L and 2D Tinkham models.[46] $H_{c2}(\theta)$ of FeSe can be fitted using the 3D model as expected [Fig. 2 (c)], yielding a reasonable small anisotropy $\gamma = 4.5$ that is only a little higher than the previously reported $\gamma$ value of 2.[47] However, as can be seen from the inset of Fig. 2 (d), $H_{c2}(\theta)$ of (Li,Fe)OHFeSe is well fitted using the 2D model, resulting in a much higher anisotropy $\gamma = 151$. We note that this anisotropy $\gamma$ is one order of magnitude larger than the result of previous work, where the $\gamma$ of (Li,Fe)OHFeSe is unsatisfactorily fitted by the 3D anisotropic G-L model.[48] Such a high anisotropy $\gamma = 151$ is comparable to that of Bi-2212 cuprates,[49] making the case that the intercalated (Li,Fe)OHFeSe superconducor is very similar to the highly anisotropic bismuth cuprate superconductors. The 3D or quasi-2D character is also evident from the line shape of the corresponding $H_{c2}(\theta)$ peak at $\theta = 90°$. The peak for the 3D anisotropic prototypical FeSe superconductor is rounded in shape [Fig. 2 (c)]. In contrast, a cusp-like peak at $\theta = 90°$ is observed for the intercalated (Li,Fe)OHFeSe superconductor [the inset of Fig. 2(d)]. This cusp-like $H_{c2}(\theta)$ peak is rare and can only be accounted for by the 2D Tinkham model, demonstrating the highly anisotropic superconductivity of (Li,Fe)OHFeSe. It should be pointed out that the thickness of the studied (Li,Fe)OHFeSe films (600-700 nm) is significantly larger than the $c$-axis SC coherence length $\xi_c = 0.24$ nm.[48] Therefore, the observed quasi-2D superconductivity is intrinsic to the (Li,Fe)OHFeSe compound, rather than induced by the sample thicknesses of the films.

The quasi-2D nature of superconductivity is also manifested in the vortex behavior under applied $c$-axis magnetic field. Actually, a noticeable difference exists between the resistive transition behavior of the FeSe and (Li,Fe)OHFeSe samples under the $c$-axis fields. As the $c$-axis magnetic field is increased, the SC transition of FeSe is suppressed in an almost parallel way, *i.e.*, the transition width is little changed [Fig. 1 (f)]. This contrasts sharply with a fan-shaped suppression of the SC transition seen in (Li,Fe)OHFeSe: While its SC transition is narrow in width at zero field, it is rather broadened at higher fields with a long resistive tail before reaching zero resistance [Fig. 1 (g)]. Although a strongly suppressed onset zero-resistance temperature is apparently due to a wide temperature range for the dissipative vortex liquid phase, the detailed mechanism underlying the notably broadened vortex liquid region by the field, which is present in the quasi-2D (Li,Fe)OHFeSe but absent in the 3D anisotropic FeSe, needs to be further elucidated. We recall that a similar long tail in the resistive transition has also been observed before in the highly anisotropic Bi-2212 cuprate superconductors.[50] The presence of the long resistive tail under the field can be understood from a dimensional crossover of the vortex structure[28] in these highly 2D systems. When the 3D collective behavior of the vortices starts to transform to a quasi-2D one above a threshold $c$-axis field $H_{2D}(T)$, the effective vortex pinning energy is dramatically reduced with the increasing field.[29] This inevitably leads to the magnetic-field-enhanced broadening of the dissipative vortex liquid region. Here the zero-temperature threshold field for the magnetically induced dimensional crossover in (Li,Fe)OHFeSe can be estimated as $H_{2D}^{0K} \sim \Phi_0/(\gamma d)^2 = 1$ kOe, with its $\gamma = 151$ and $d \sim 9.3$ Å.

Therefore, we further investigate the vortex dynamics to experimentally determine the dimensional crossover field $H_{2D}(T)$ of (Li,Fe)OHFeSe by the magnetic relaxation measurements on the single-crystal samples. The studied (Li,Fe)OHFeSe single crystals also exhibit a sharp SC diamagnetic transition at $H = 1$ Oe (with a narrow transition width $\Delta T \sim 1$ K between the 0 % and 100 % diamagnetic shielding), indicating their high sample quality. And the $T_c = 42.0$ K, determined from the onset temperature of the diamagnetic signals [Fig. 3 (a)], is also in good agreement with that of the (Li,Fe)OHFeSe films by the resistive transition. In Fig. 3(b), we show the representative relaxation of magnetizations in the double-logarithmic scales, from which the normalized magnetic relaxation rate $S$ can be extracted by fitting the line slopes. The high fitting index, $R^2 > 99.9\%$, ensures the reliability of the obtained $S$ values. The magnetic field dependence of $S$ and $U^*$ ($=T/S$) at various $T < T_c$ are displayed in Figs. 3(c) and (d), respectively. The change in the $U^*$ value reflects the change in the effective vortex pinning energy.[43] Therefore, we define the crossover field $H_{2D}$ at a given $T$ as the intersection point of the two distinct behavior of $U^*$ vs $H$, as marked by the arrows in Fig. 3(d). While $U^*$ vs $H$ generally shows a plateau below $H_{2D}$ (except for a slow rise only at $T = 40$ K), it starts to drop above $H_{2D}$. This change in $U^*$ behavior demonstrates the existence of two distinct vortex dynamics regimes divided by $H_{2D}$, with a reduced effective vortex pinning energy at $H > H_{2D}$. The measured $H_{2D}(T)$ is about 0.04 kOe at $T = 40$ K, it progressively increases with decreasing $T$ as expected and eventually get quite close to the theoretical zero-temperature $H_{2D}^{0K} \sim 1$ kOe at the lowest measuring temperature $T$ = 2 K. This consistency between the experiment and theory indicates that the measured $H_{2D}(T)$ values are intimately connected with the vortex dimensional crossover behavior of the strongly layered (Li,Fe)OHFeSe. We note that such a vortex dimensional crossover of (Li,Fe)OHFeSe is very similar to the case of Bi-2212 cuprates.[26,27]

On the other hand, it has been shown that, in strongly layered superconductors, the applied in-plane electrical current can also induce quasi-2D vortex-antivortex pair excitations, even though no magnetic field is applied.[28,51] The corresponding transition can be explained by the BKT theory proposed for a 2D system.[52,53] We indeed observed a BKT-like transition in (Li,Fe)OHFeSe. As seen in Fig. 4 (a), its resistive transition has the characteristic of a BKT type one, obeying the Halperin-Nelson scale [dln$R$/d$T$]$^{-2/3}$ near the BKT transition temperature $T_{BKT}$.[54,55] Furthermore, in Fig. 4 (b) we show the voltage-current (*V-I*) curves measured at different temperatures. The exponent $\alpha$, fitted from the power law $V \propto I^\alpha$ [Fig. 4 (b)], jumps with decreasing temperature and exceeds the value of 3 at $T_{BKT}$ [Fig. 4 (c)]. This provides another evidence for a BKT type transition.[54] The BKT transition temperature obtained from the resistance scaling, $T_{BKT}$ = 42.1 K [Fig. 4 (a)], is very close to that given at $\alpha(T_{BKT})$ = 3, $T_{BKT}$ = 42.0 K [Fig. 4 (c)], and both of them are slightly below the fitted mean-field superconducting $T_c'$ = 43.2 K [inset of Fig. 4 (a)] as expected. These evidences support the existence of a BKT-like transition in (Li,Fe)OHFeSe, again similar to its Bi-2212 counterpart.[51]

By combining together all the experimental and fitting results as well as the theoretical prediction, we can establish a vortex phase diagram of the *c*-axis field vs temperature for (Li,Fe)OHFeSe superconductor, shown in Fig. 5. The irreversibility field $H_{irr}(T)$ and upper critical field $H_{c2}(T)$ extracted at 0.1 % and 50 % $R_n$, respectively, normally separate the vortex phase diagram into three regions: The normal state, the vortex liquid, and the vortex solid. Additionally, here a new line (red dashed line) of the vortex dimensional crossover field $H_{2D}(T)$ is added within the vortex solid region. The field-induced vortex dimensional crossover at zero temperature is indicated in the vertical axis by the theoretical $H_{2D}^{0K}$ ~ 1 kOe. Note that the temperature range of the dissipative vortex liquid phase is significantly broadened by the magnetic fields above the threshold $H_{2D}^{0K}$. The quasi-2D nature of the dissipative vortex liquid state has been demonstrated by the 2D Tinkham fitting of $H_{c2}(\theta)$ [Fig. 2 (d)]. For the case at zero external field, the BKT-like transition ($T_{BKT}$ = 42.1 K) occurring just below the mean-field SC transition ($T_c'$ = 43.2 K) is indicated in the horizontal axis. The presence of the $H_{2D}(T)$ line indicates that the long-range interlayer coupling in the 3D ground state at zero temperature and field is transformed to the Josephson-type one between the quasi-2D pancake vortices[29,31] with increasing temperature, even without applying the magnetic field. Therefore, the superconducting transition at the zero field is also quasi-2D in nature, consistent with the observation of the BKT-like transition. Thus, this $H$-$T$ phase diagram elucidates the emergent quasi-2D vortex-liquid state near the superconducting transition, and the subsequent vortex-solid dimensional crossover at $H_{2D}(T)$ from the quasi-2D pancake to the 3D vortex state that exists at lower temperatures and fields.

**Further discussions**

First of all, the quasi-2D nature of the vortex liquid and solid states is essential for the highly anisotropic superconductivity appearing in (Li,Fe)OHFeSe. On the one hand, previous studies of scanning tunneling microscopy (STM) have found the evidence for the topological superconductivity in the (Li,Fe)OHFeSe epitaxial films,[23,56,57] and pointed out that the interlayer coupling plays an important role in the topological superconductivity.[24] The STM experiments[23] were conducted at a low temperature $T$ = 4.2 K and a $c$-axis magnetic field $H$ = 100 kOe far above the threshold $H_{2D}$ identified here; namely, within the quasi-2D pancake-vortex regime of the phase diagram in Fig. 5. On the other hand, the phase diagram indicates that the highly anisotropic superconductivity at the zero field is also closely related to the quasi-2D vortices. Moreover, it is important to mention that the quasi-2D magnetic fluctuations and electrical transport have been shown to persist into the normal state up to a temperature well above $T_c$ in our earlier study of (Li,Fe)OHFeSe,[33] and the quasi-2D transport has also been observed in Bi-2212 cuprates.[58] Therefore, based on the present vortex phase diagram (Fig. 5) and the previous related results, we further highlight that the interlayer coupling between the emergent quasi-2D vortices plays a significant role in the appearance of the highly anisotropic high-$T_c$ and topological superconductivity in (Li,Fe)OHFeSe.

We have also measured the magnetic hysteresis loops (MHLs) on the FeSe and (Li,Fe)OHFeSe single crystals, the results are shown in supplemental Figs. S1 (a) and (c), respectively. The symmetric loops indicate that the bulk pinning is dominating in both FeSe and (Li,Fe)OHFeSe, consistent with the previous reports.[59,60] The field-dependent critical current density $J_c(H)$ is calculated from the MHLs using the Bean critical model[61] for FeSe [Fig. S1 (b)] and (Li,Fe)OHFeSe [Fig. S1 (d)]. The self-field $J_c$ at $T$ = 2 K increases from 0.045 MA/cm$^{-2}$ in FeSe to 1.9 MA/cm$^{-2}$ in (Li,Fe)OHFeSe, indicating a high $J_c$ value achievable in high-$T_c$ (Li,Fe)OHFeSe.[48,62] We note that, distinct from a monotonic $J_c$ variation of FeSe, $J_c$ of (Li,Fe)OHFeSe shows a local maximum $J_{c,max}(H, T)$ at a given $H$ and $T$ [see inset of Fig. S1 (d)]. The $J_{c,max}$ observed for (Li,Fe)OHFeSe is reminiscent of a well-known fishtail effect, which corresponds to the peak appearing in its MHLs [inset of Fig. S1 (c)], though the origin of fishtail effect is debatable.[63] Interestingly, however, here we find that the position of $J_{c,max}(H, T)$ generally follows the dimensional crossover field $H_{2D}(T)$ in the $H$-$T$ phase diagram [Fig. 5]. This indicates that $J_{c,max}$ is more or less coupled to the vortex dimensional crossover, as also proposed before for Bi-2212 cuprates.[27]

In addition, the Josephson-type interlayer coupling between the quasi-2D pancake vortices is a prerequisite for the intrinsic Josephson junctions (IJJs). The IJJ systems have attracted considerable interest in practical applications like the terahertz sources.[64] So far, the IJJs have been confirmed in the single-band cuprate superconductors, mainly in the Bi-2212 system.[31,64] The IJJs in multi-band superconductor have only been reported for the iron-arsenide $(V_2Sr_4O_6)Fe_2As_2$ in a micron-sized device prepared by a focused ion beam method.[65] However, no further study of the IJJs is reported in the multi-band iron-based superconductor, presumably due to the difficulty in growing large $(V_2Sr_4O_6)Fe_2As_2$ single crystals. Our identification of the quasi-2D pancake vortices in (Li,Fe)OHFeSe signifies a new chance to study the IJJ phenomena in the iron-based compound, since the large single crystals and epitaxial films of (Li,Fe)OHFeSe are experimentally available now.

**Summary**

We reveal the quasi two-dimensional nature of the superconductivity in the intercalated (Li,Fe)OHFeSe, and establish a new vortex phase diagram. We also find that the high-$T_c$ iron selenide (Li,Fe)OHFeSe resembles the high-$T_c$ bismuth cuprate $Bi_2Sr_2CaCu_2O_{8+\delta}$ in all respects of the two-dimensionality effect, including the highly anisotropic superconductivity with $\gamma =$ 151, the associated quasi-2D vortex liquid and solid states, the vortex-solid dimensional crossover at a threshold $H_{2D}(T)$, the magnetic-field-broadened vortex liquid region, the BKT-like transition at zero external magnetic field, as well as the quasi-2D properties persisting into the normal state.

Therefore, our results provide important evidence for the common quasi-2D physics of high-$T_c$ superconductivity in the multi-band iron-based (Li,Fe)OHFeSe and single-band copper-based $Bi_2Sr_2CaCu_2O_{8+\delta}$ compounds. Particularly in (Li,Fe)OHFeSe, the interlayer coupling between the emergent quasi-2D vortices plays a significant role in the appearance of both the high-$T_c$ and topological superconductivity. In addition, our results also suggest that the (Li,Fe)OHFeSe superconducting material is promising in potential applications of the quasi-2D pancake vortices.

## Acknowledgments

We thank Mingwei Ma for technical assistance in preparing the FeSe single crystals. This work was supported by the National Natural Science Foundation of China (Grant Nos. 12061131005, 11834016, and 11888101), and the Strategic Priority Research Program of Chinese Academy of Sciences (Grant Nos. XDB33010200 and XDB25000000).

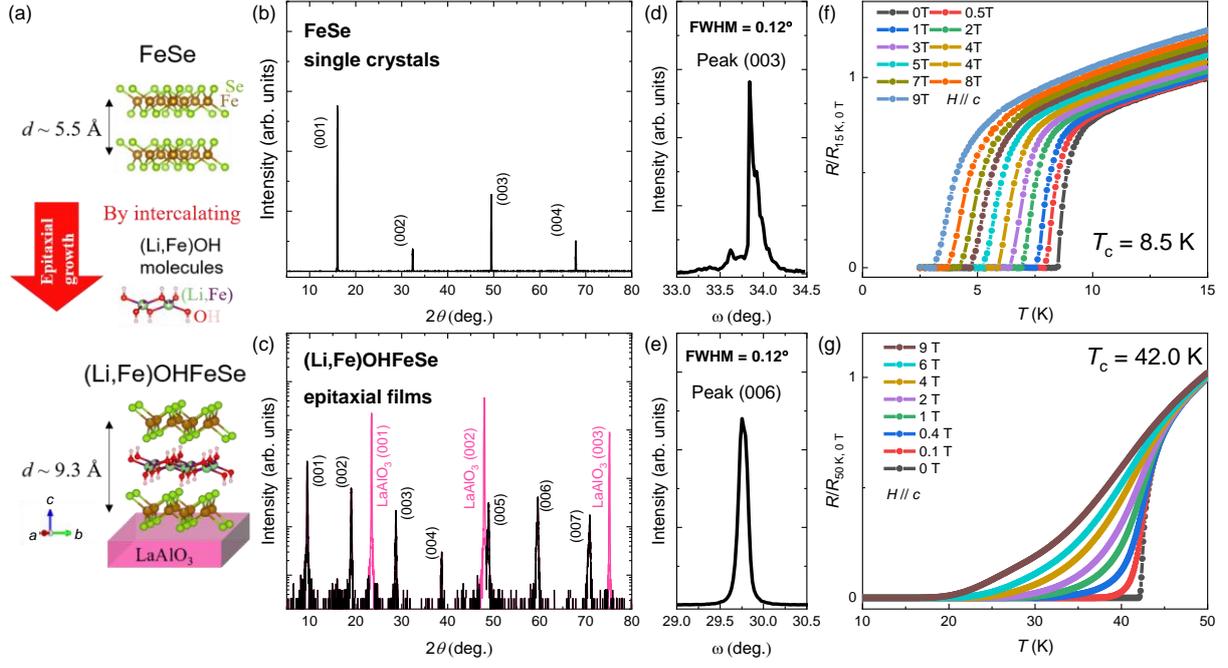

**Fig. 1. The crystal structure, XRD characterization, and superconducting transition behavior of FeSe and (Li,Fe)OHFeSe.** (a) Schematic illustrations of the crystal structures of FeSe single crystals and (Li,Fe)OHFeSe epitaxial films grown on LaAlO$_3$ substrates. The strongly layered structure of (Li,Fe)OHFeSe ($d \sim 9.3$ Å) is achieved by intercalating (Li,Fe)OH molecules into the bulk FeSe structure ($d \sim 5.5$ Å). (b) and (c) The (00$l$) XRD patterns of the samples. (d) and (e) X-ray rocking curves for the (003) peak of FeSe and (006) peak of (Li,Fe)OHFeSe, respectively. (f) and (g) Temperature dependence of the normalized resistance near $T_c$ under $c$-axis fields up to 9 T for FeSe single crystals ($T_c = 8.5$ K) and (Li,Fe)OHFeSe epitaxy films ($T_c = 42.0$ K), respectively.

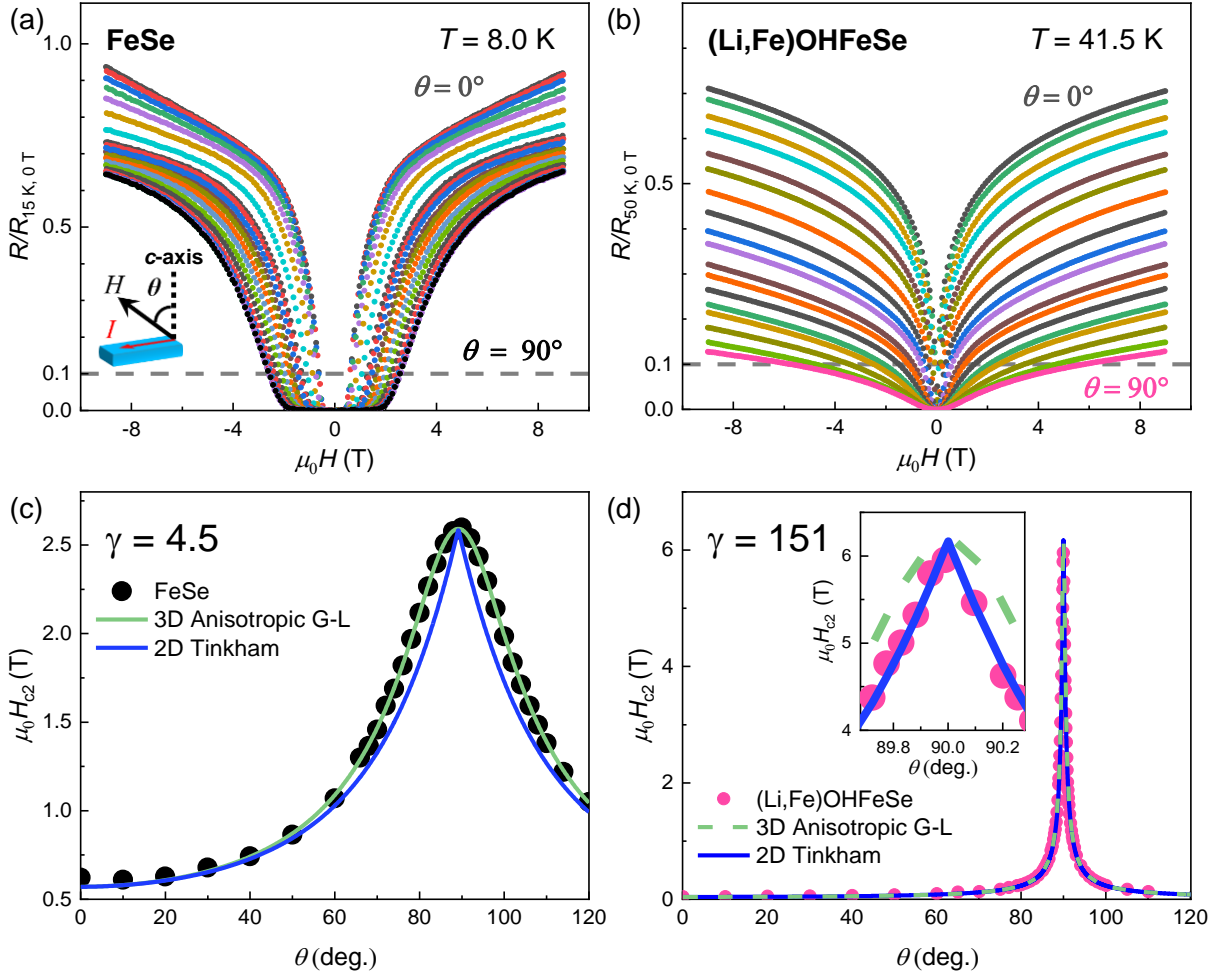

**Fig. 2. The 3D and 2D model fittings of the superconductivity for FeSe and (Li,Fe)OHFeSe, respectively.** (a) and (b) Magnetic field dependence of the normalized resistance measured at different tilting angles for the FeSe single crystals (at $T = 8.0$ K) and (Li,Fe)OHFeSe epitaxial films (at $T = 41.5$ K), respectively. (c) and (d) Angular dependence of the upper critical field $\mu_0 H_{c2}$ extracted at 10 % $R_n$ for the FeSe and (Li,Fe)OHFeSe samples, respectively. The olive and blue curves represent the fittings by the 3D anisotropic mass model, $\left(H_{c2}(\theta)\cos\theta/H_{c2}^c\right)^2 + \left(H_{c2}(\theta)\sin\theta/H_{c2}^{ab}\right)^2 = 1$, and the 2D Tinkham formula, $\left|H_{c2}(\theta)\cos\theta/H_{c2}^c\right| + \left(H_{c2}(\theta)\sin\theta/H_{c2}^{ab}\right)^2 = 1$, respectively.

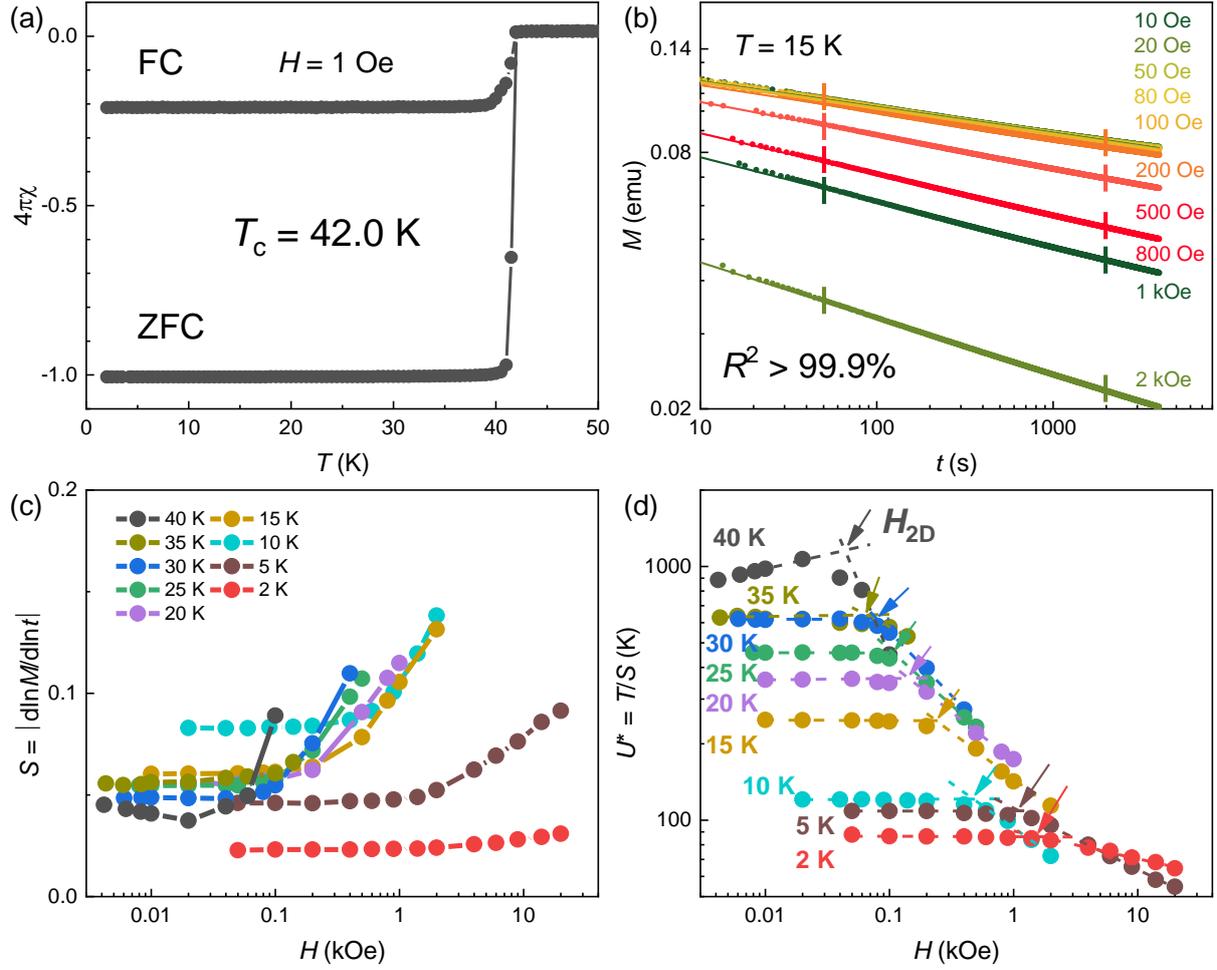

**Fig. 3. Magnetization measurements of (Li,Fe)OHFeSe single cryastals.** (a) Temperature dependence of the DC magnetic susceptibility measured at $H = 1$ Oe and $T \leq 50$ K with the zero-field-cooled (ZFC) and field-cooled (FC) modes. (b) The representative time $t$ dependence of the magnetization measured at different fields and $T = 15$ K. (c) Magnetic field dependence of the fitted relaxation rate $S$ at different $T \leq 40$ K. (d) Double-logarithmic plots of the effective pinning energy $U^*$ ($= T/S$) as functions of magnetic field at different temperatures. The vortex dimensional crossover fields ($H_{2D}$'s) are obtained at the intersection points (marked by the arrows), above which the effective pinning energy starts to drop.

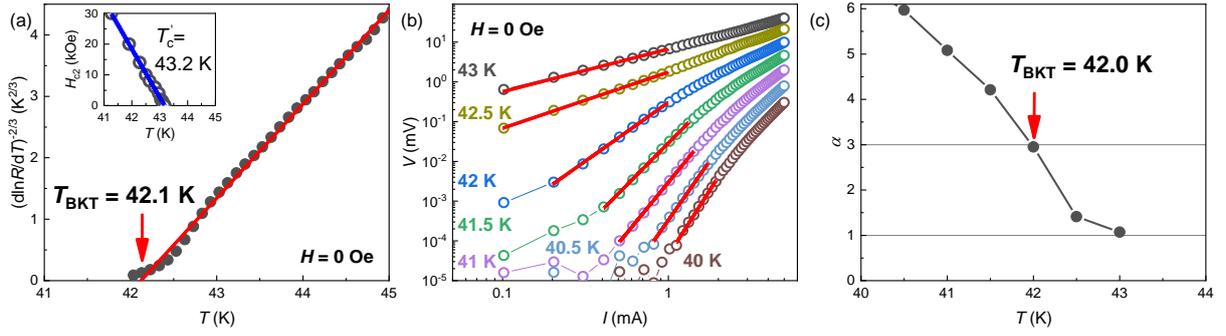

**Fig. 4. The BKT-like transition in (Li,Fe)OHFeSe expitaxial films.** (a) The resistive transition of (Li,Fe)OHFeSe is plotted as $[d\ln R/dT]^{-2/3}$ vs $T$, with the BKT transition temperature obtained as $T_{BKT} \sim 42.1$ K by the linear fitting (the red line). The inset shows the linear temperature dependence of $H_{c2}$ and the mean-field superconducting $T_c' = 43.2$ K obtained by the G-L fitting $H_{c2}(T) \propto 1 - T/T_c'$ (the blue line). (b) Voltage-current ($V$-$I$) curves of (Li,Fe)OHFeSe measured at different temperatures. The red lines are the power-law ($V \propto I^\alpha$) fits to the curves. (c) Temperature dependence of the power-law fitted exponent $\alpha$. The corresponding $T_{BKT} \sim 42.0$ K is obtained at $\alpha(T_{BKT}) = 3$.

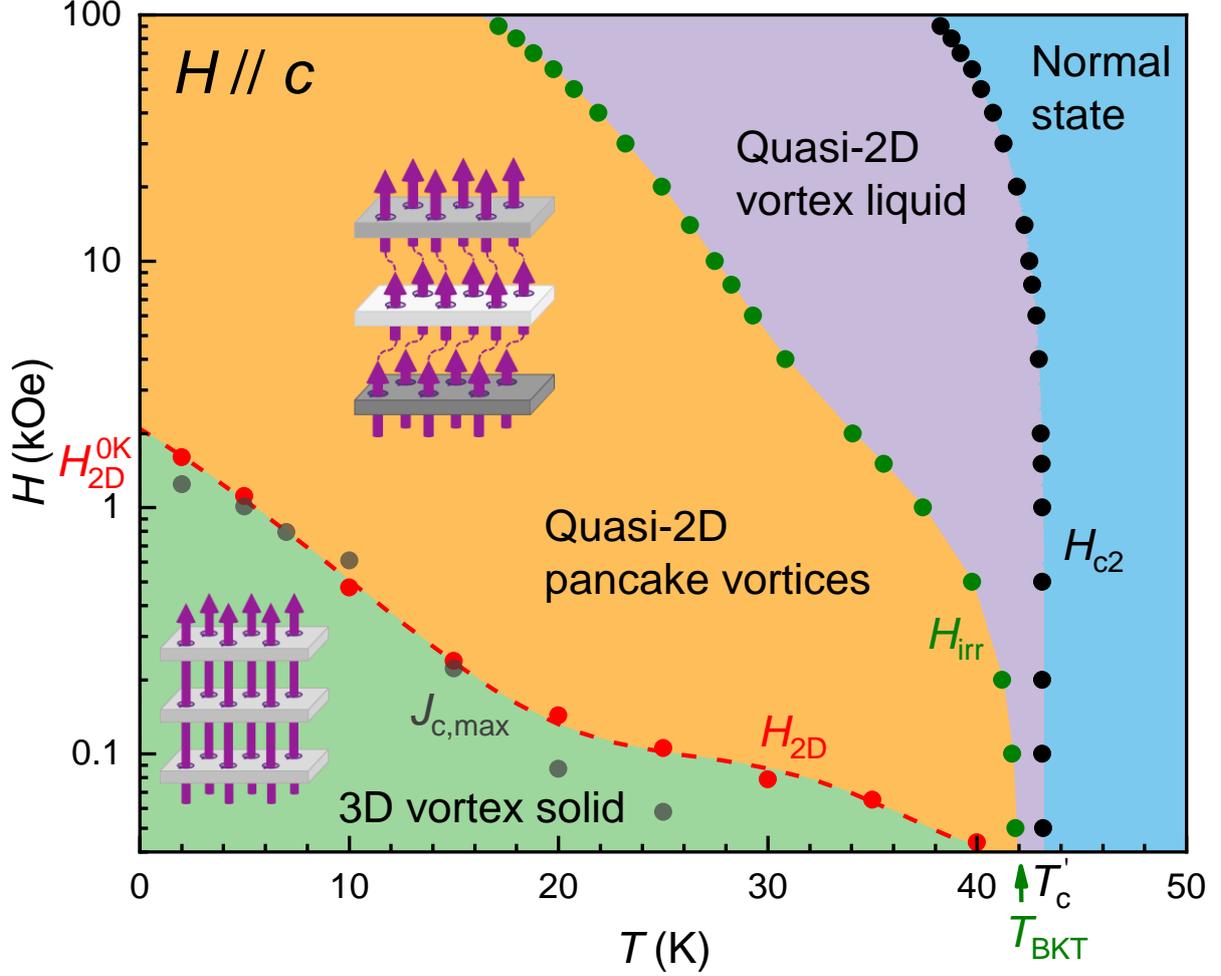

**Fig. 5. *H-T* vortex phase diagram of (Li,Fe)OHFeSe superconductor ($T_c$ = 42.0 K).** The phase diagram puts stress on the quasi-2D nature of the vortex liquid and solid states, as well as the new red dashed line of the dimensional crossover field $H_{2D}(T)$ (red circles). The routine irreversibility field $H_{irr}(T)$ and upper critical field $H_{c2}(T)$ are represented by the greeen and black circles, respectively. The quasi-2D (pancake) and 3D vortex solid states are schematically illustrated by the corresponding insets. The gray circles represent the positions of $J_{c,max}(H, T)$. The zero-temperature threshold $H_{2D}^{0K}$ ~ 1 kOe is indicated in the vertical axis, and the zero-field BKT $T_{BKT}$ = 42.1 K and mean-field SC $T_c'$ = 43.2 K are indicated in the horizontal axis

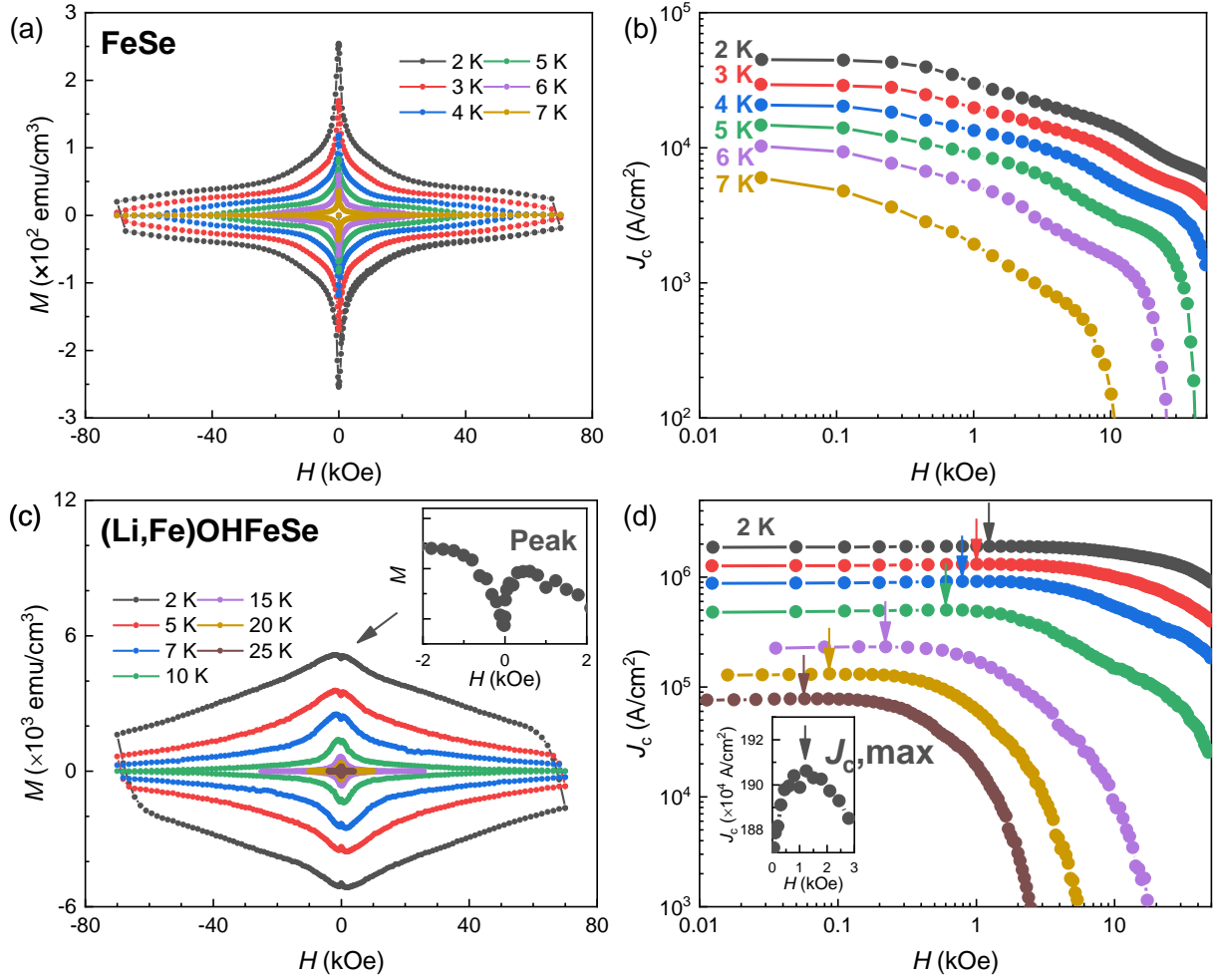

**Fig. S1. Magnetic hysteresis loops (MHLs) of FeSe and (Li,Fe)OHFeSe single cryastals.** (a) The MHLs of FeSe measured at temperatures from 2 to 7 K. (b) Magnetic field dependence of $J_c$ claculated from the MHLs in (a). (c) The MHLs of (Li,Fe)OHFeSe measured at temperatures from 2 to 25 K. The inset zooms in the MHL at 2 K to show the peak feature. (d) Magnetic field dependence of $J_c$ claculated from the MHLs in (c). The $J_c$ shows a local maximum $J_{c,max}$ in the temperature-dependent field at each given $T$, as marked by the arrows. The inset zooms in the $J_c(H)$ at 2 K to show its $J_{c,max}$.

.